\documentclass[%
reprint,
superscriptaddress,
frontmatterverbose,
showpacs,
preprintnumbers,
nofootinbib,
nobibnotes,
amsmath,amssymb,
aps,
prc,
]{revtex4-1}
\usepackage{graphicx}
\usepackage{dcolumn}
\usepackage{bm}
\usepackage{CJK}
\usepackage[T1]{fontenc}
\def\ket#1{\left|{#1}\right\rangle}

\def\brakket#1#2#3{\left\langle{#1}\middle|{#2}\middle|{#3}\right\rangle}
\def\ve#1{{\bm{#1}}}
\def\nuc#1#2#3{{}^{#2}_{#3}\mathrm{#1}}
\def\urm#1{\scriptstyle{\text{\textrm{\textmd{\textup{#1}}}}}}

\def\avr#1{\left\langle{#1}\right\rangle}
\def\Nabla{\bm{\nabla}}
\def\ca#1{{\mathcal{#1}}}
\let\temp\epsilon
\let\epsilon\varepsilon
\let\varepsilon\temp
\let\temp\relax
\begin{document}
%
\begin{CJK*}{UTF8}{}
  \preprint{RIKEN-QHP-473}
  \preprint{RIKEN-iTHEMS-Report-20}
  \title{
    Effects of finite nucleon size, vacuum polarization, and electromagnetic spin-orbit interaction on nuclear binding energies and radii in spherical nuclei}
  \author{Tomoya Naito (\CJKfamily{min}{内藤智也})}
  \affiliation{Department of Physics, Graduate School of Science, The University of Tokyo,
    Tokyo 113-0033, Japan}
  \affiliation{RIKEN Nishina Center, Wako 351-0198, Japan}
  \author{Xavier Roca-Maza}
  \affiliation{Dipartimento di Fisica, Universit\`{a} degli Studi di Milano,
    Via Celoria 16, 20133 Milano, Italy}
  \affiliation{INFN, Sezione di Milano,
    Via Celoria 16, 20133 Milano, Italy}
  \author{Gianluca Col\`{o}}
  \affiliation{Dipartimento di Fisica, Universit\`{a} degli Studi di Milano,
    Via Celoria 16, 20133 Milano, Italy}
  \affiliation{INFN, Sezione di Milano,
    Via Celoria 16, 20133 Milano, Italy}
  \author{Haozhao Liang (\CJKfamily{gbsn}{梁豪兆})}
  \affiliation{RIKEN Nishina Center, Wako 351-0198, Japan}
  \affiliation{Department of Physics, Graduate School of Science, The University of Tokyo,
    Tokyo 113-0033, Japan}
  \date{\today}
  \begin{abstract}
    The electromagnetic effects of the finite size of the nucleon
    are implemented self-consistently on top of the Skyrme Hartree-Fock calculation,
    where the electric form factors of both protons and neutrons are considered.
    Furthermore, the vacuum polarization and the electromagnetic spin-orbit interaction are taken into account.
    The self-consistent finite-size effects give a different Coulomb potential from the conventional one and affect the neutrons as well.
    The contribution of the finite-size effects to the total energy reaches $ 7 \, \mathrm{MeV} $ in $ {}^{208} \mathrm{Pb} $.
    The vacuum polarization and the electromagnetic spin-orbit interaction are also non-negligible, especially in the heavy nuclei.
    These effects provide a comparable contribution of the total energy to that of the isospin symmetry-breaking terms of the nuclear interaction.
    The mirror nuclei mass difference in $ {}^{48} \mathrm{Ca} $--$ {}^{48} \mathrm{Ni} $ is also studied,
    and its value is improved by approximately one order of magnitude.
  \end{abstract}
  \maketitle
\end{CJK*}
%
\section{Introduction}
\par
Atomic nuclei are quantum many-body systems composed of protons and neutrons,
which interact with each other by the nuclear and Coulomb interactions.
It is known that the nuclear interaction is much stronger than the Coulomb interaction,
and thus the main contribution to the nuclear properties comes from the nuclear interaction.
Nevertheless,
the study of the Coulomb effects on the properties is also important, 
since the Coulomb interaction and the isospin symmetry-breaking (ISB) terms of the nuclear interaction are entangled to each other
in some particular nuclear properties,
such as the superallowed $ \beta $ decay 
\cite{Hardy2015Phys.Rev.C91_025501, Liang2009Phys.Rev.C79_064316}, 
the energy difference of the mirror nuclei 
and its Nolen-Schiffer anomaly
\cite{Okamoto1964Phys.Lett.11_150, Nolen1969Annu.Rev.Nucl.Sci.19_471, Saito1994Phys.Lett.B335_17, Shahnas1994Phys.Rev.C50_2346, Meisner2008Eur.Phys.J.A36_37, Menezes2009Eur.Phys.J.A42_97, Dong2018Phys.Rev.C97_021301}, 
and the isobaric analog states (IASs)
\cite{Jaenecke1965Nucl.Phys.73_97, Shlomo1978Rep.Prog.Phys.41_957, Roca-Maza2018Phys.Rev.Lett.120_202501}.
Recently, to study such ISB effects,
a parametrization of the ISB terms of the nuclear interaction that can reproduce the mirror and triplet displacement energies on top of a Skyrme functional was proposed
\cite{Baczyk2018Phys.Lett.B778_178, Baczyk2019J.Phys.G46_03LT01}.
From these and other works, it is evident that medium effects are relevant for those ISB terms, and the approximation of using the bare coupling constants is not adequate.
Our current work does not address the important question of understanding the relationship between bare and effective ISB forces, but rather it focuses on a complementary and yet relevant issue:
Our scope is to study the other Coulomb-related terms in DFT more precisely than has been done so far.
\par
The density functional theory (DFT) 
\cite{Hohenberg1964Phys.Rev.136_B864, Kohn1965Phys.Rev.140_A1133}
is one of the powerful and widely used methods to solve the quantum many-body problem.
In the DFT for nuclear physics, the ground-state energy is usually given by
\begin{equation}
  \label{eq:Egs}
  E_{\urm{gs}}
  =
  T_0 
  +
  E_{\urm{nucl}} \left[ \rho_p, \rho_n \right]
  +
  E_{\urm{Cd}} \left[ \rho_{\urm{ch}} \right]
  +
  E_{\urm{Cx}} \left[ \rho_{\urm{ch}} \right],
\end{equation}
where
$ T_0 $ is the Kohn-Sham kinetic energy,
and
$ E_{\urm{nucl}} $, $ E_{\urm{Cd}} $, and $ E_{\urm{Cx}} $ are the energy density functionals (EDFs) of nuclear, Coulomb direct, and Coulomb exchange parts, respectively.
Here, $ \rho_p $ and $ \rho_n $ are the ground-state density distributions of protons and neutrons, respectively,
and $ \rho_{\urm{ch}} $ is the charge density distribution
\cite{Bender2003Rev.Mod.Phys.75_121, Nakatsukasa2016Rev.Mod.Phys.88_045004, Roca-Maza2018Prog.Part.Nucl.Phys.101_96}.
Since the nuclear EDF $ E_{\urm{nucl}} $ is usually fit to the experimental data with certain Ans\"{a}tze,
such as the Skyrme type 
\cite{Vautherin1972Phys.Rev.C5_626},
the Gogny type
\cite{Berger1991Comput.Phys.Commun.63_365},
and the relativistic one 
\cite{Meng2006Prog.Part.Nucl.Phys.57_470, Liang2015Phys.Rep.570_1},
because of the missing knowledge of the nuclear force in medium, 
$ E_{\urm{nucl}} $ includes the Coulomb correlation
as well as the nuclear one implicitly.
In contrast, 
the Coulomb EDFs, $ E_{\urm{Cd}} $ and $ E_{\urm{Cx}} $, are deduced theoretically as accurately as possible,
since the Coulomb interaction is well known.
\par
Although the Coulomb EDFs can be given fully theoretically,
the Hartree-Fock-Slater \cite{Dirac1930Proc.Camb.Phil.Soc.26_376, Slater1951Phys.Rev.81_385} 
or even Hartree approximation has been widely used
\cite{Bender2003Rev.Mod.Phys.75_121, Nakatsukasa2016Rev.Mod.Phys.88_045004}.
Recently, effects of the exact-Fock treatment \cite{Gu2013Phys.Rev.C87_041301, VanGiai2014Phys.Scr.89_054008, Roca-Maza2016Phys.Rev.C94_044313} 
and those beyond the Hartree-Fock-Slater approximation, the so-called generalized gradient approximation (GGA) 
\cite{Naito2018Phys.Rev.C97_044319, Naito2019Phys.Rev.C99_024309},
were discussed in the context of the nuclear DFT
to achieve more accurate evaluation of the Coulomb contribution to the total energy.
\par
Moreover, the role of the Coulomb functional has several open questions.
For example, it was suggested that the Coulomb exchange term is almost canceled out by the ISB terms of the nuclear force, 
and thus the Skyrme functionals fit without the Coulomb exchange term reproduce the masses better than those with it 
\cite{AlexBrown1998Phys.Rev.C58_220, Chamel2009Phys.Rev.C80_065804}.
The effective charge in the nuclear DFT has also been discussed
\cite{Niu2013Phys.Rev.C87_037301, Dong2019Nucl.Phys.A983_133}.
Recently, Dong \textit{et al\/.}~\cite{Dong2019Nucl.Phys.A983_133} showed that, 
in the Hartree-Fock-Slater approximation,
i.e.,~the local density approximation (LDA) for the Coulomb exchange term,
introducing the effective coupling constant $ e_0^2 = e^2 \left( 1 + a Z^{-2/3} \right) $
reproduces the isobaric multiplet mass equation well.
They also discussed that this effective coupling constant includes all the possible electromagnetic (EM) contribution,
such as the difference between the LDA and the exact-Fock, the finite-size effects, and the vacuum polarization.
\par
The Coulomb EDFs $ E_{\urm{Cd}} $ and $ E_{\urm{Cx}} $ are, in principle, written in terms of the charge density $ \rho_{\urm{ch}} $ \cite{Bulgac1996Nucl.Phys.A601_103},
because the Coulomb interaction affects the charge itself instead of the point protons. 
Nevertheless, the protons and neutrons are assumed to be point particles, 
i.e.,~$ \rho_{\urm{ch}} \equiv \rho_p $ is assumed 
(hereafter, this approximation is called the point-particle approximation),
in most of the self-consistent nuclear DFT.
Only a few works,
e.g.,~Refs.~\cite{Chamel2009Phys.Rev.C80_065804, Auerbach1972Rev.Mod.Phys.44_48, Roca-Maza2018Phys.Rev.Lett.120_202501}, 
considered the difference between $ \rho_{\urm{ch}} $ and $ \rho_p $.
\footnote{
  Note that the nuclear interaction is constructed as the nucleons are point particles,
  and thus the finite-size effects need not be considered in $ E_{\urm{nucl}} $.}
It was shown that the finite-size effects of nucleons,
i.e.,~the difference between $ \rho_{\urm{ch}} $ and $ \rho_p $,
are non-negligible in the energy of the isobaric analog state $ E_{\urm{IAS}} $ \cite{Auerbach1972Rev.Mod.Phys.44_48, Roca-Maza2018Phys.Rev.Lett.120_202501}.
Moreover, it was shown in Refs.~\cite{Auerbach1972Rev.Mod.Phys.44_48, Roca-Maza2018Phys.Rev.Lett.120_202501} that the vacuum polarization is also non-negligible in $ E_{\urm{IAS}} $.
\par
Although these discussions related to the Coulomb interaction have been done for decades as mentioned above,
deeper and self-consistent analysis is still desired.
Thus, in this paper, the finite-size effects of nucleons are implemented to the self-consistent steps of the Skyrme Hartree-Fock calculation.
The electric form factors of both protons and neutrons are considered.
Also, other possible EM contributions, 
i.e.,~the vacuum polarization and the EM spin-orbit interaction, are considered.
\par
This paper is organized as follows:
First, the theoretical framework is given in Sec.~\ref{sec:theo}.
Second, the simple estimation of systematic behavior of each effect is discussed in Sec.~\ref{sec:simple}.
Then, the Skyrme Hartree-Fock calculation is performed to discuss systematics and to compare experimental data of mirror nuclei mass difference in Sec.~\ref{sec:calc}.
Finally, the conclusion and future perspectives are given in Sec.~\ref{sec:conc}.
%
\section{Theoretical Framework}
\label{sec:theo}
\par
In this section, 
the theoretical frameworks of the finite-size effects, the vacuum polarization, and the EM spin-orbit interaction are shown.
The Coulomb functional is accordingly composed of three terms:
the Coulomb direct term $ E_{\urm{Cd}} $,
the Coulomb exchange term $ E_{\urm{Cx}} $,
and 
the vacuum polarization term $ E_{\urm{VP}} $.
Equation \eqref{eq:Egs} is now modified as
\begin{equation}
  E_{\urm{gs}}
  =
  T_0 
  +
  E_{\urm{nucl}} \left[ \rho_p, \rho_n \right]
  +
  E_{\urm{Cd}} \left[ \rho_{\urm{ch}} \right]
  +
  E_{\urm{Cx}} \left[ \rho_{\urm{ch}} \right]
  +
  E_{\urm{VP}} \left[ \rho_{\urm{ch}} \right],
\end{equation}
and correspondingly the effective potential is
\begin{equation}
  V_{\urm{eff} \tau} \left( \ve{r} \right)
  =
  V_{\urm{nucl} \tau} \left( \ve{r} \right)
  +
  V_{\urm{Cd} \tau} \left( \ve{r} \right)
  +
  V_{\urm{Cx} \tau} \left( \ve{r} \right)
  +
  V_{\urm{VP} \tau} \left( \ve{r} \right),
\end{equation}
where $ \tau = p $ ($ n $) for protons (neutrons),
and $ V_{\urm{nucl} \tau} $, $ V_{\urm{Cd} \tau} $, $ V_{\urm{Cx} \tau} $, and $ V_{\urm{VP} \tau} $ are
the effective potentials coming from the nuclear force, Coulomb direct and exchange terms, and the vacuum polarization, respectively.
\par
In $ E_{\urm{Cd}} $ and $ E_{\urm{Cx}} $, the finite-size effects of nucleons are considered,
where the Coulomb direct term $ E_{\urm{Cd}} $ holds the conventional form 
\begin{equation}
  \label{eq:ECd} 
  E_{\urm{Cd}} \left[ \rho_{\urm{ch}} \right]
  =
  \frac{e^2}{2}
  \iint
  \frac{
    \rho_{\urm{ch}} \left( \ve{r} \right) \, 
    \rho_{\urm{ch}} \left( \ve{r}' \right)}
  {\left| \ve{r} - \ve{r}' \right|}
  \, d \ve{r} \, d \ve{r}'.
\end{equation}
\par
The EM spin-orbit interaction is considered perturbatively.
\subsection{Coulomb Exchange Functional in Generalized Gradient Approximation}
\par
In principle, the nonlocal form of the Fock term
\begin{equation}
  E_{\urm{F}}
  =
  - \frac{e^2}{2}
  \sum_{i, j}
  \iint
  \frac{
    \psi_i^* \left( \ve{r} \right)
    \psi_j^* \left( \ve{r}' \right)
    \psi_i \left( \ve{r}' \right)
    \psi_j \left( \ve{r} \right)}{\left| \ve{r} - \ve{r}' \right|}
  \, d \ve{r} \, d \ve{r}'
\end{equation}
should be used for the Coulomb exchange term in the point-particle approximation,
where $ \psi_i $ is the single-particle wave function of protons.
The GGA for the Coulomb exchange term was proposed in the context of the nuclear DFT \cite{Naito2018Phys.Rev.C97_044319,Naito2019Phys.Rev.C99_024309}.
The GGA functional reproduces the Coulomb exact-Fock energy within $ 100 \, \mathrm{keV} $ error.
In this paper, the GGA Coulomb exchange functional is used instead of the exact-Fock term,
because writing the functional in terms of density has the advantage of considering the finite-size effects, which we discuss later.
\par
The GGA Coulomb exchange functional is written as \cite{Perdew1996Phys.Rev.Lett.77_3865}
\begin{equation}
  \label{eq:ECx} 
  E_{\urm{Cx}} \left[ \rho_{\urm{ch}} \right]
  =
  \int
  \epsilon_{\urm{Cx}}^{\urm{LDA}} \left( \rho_{\urm{ch}} \left( \ve{r} \right) \right) \,
  F \left( s \left( \ve{r} \right) \right)
  \rho_{\urm{ch}} \left( \ve{r} \right) 
  \, d \ve{r} ,
\end{equation}
where $ \epsilon_{\urm{Cx}}^{\urm{LDA}} $ is the LDA exchange energy density
\begin{equation}
  \epsilon_{\urm{Cx}}^{\urm{LDA}} \left( \rho_{\urm{ch}} \right)
  =
  -
  \frac{3 e^2}{4}
  \left( \frac{3}{\pi} \right)^{1/3}
  \rho_{\urm{ch}}^{1/3},
\end{equation}
$ s $ is the dimensionless density gradient
\begin{equation}
  s
  =
  \frac{\left| \Nabla \rho_{\urm{ch}} \right|}
  {2 k_{\urm{F}} \rho_{\urm{ch}}},
  \qquad
  k_{\urm{F}}
  =
  \left( 3 \pi^2 \rho_{\urm{ch}} \right)^{1/3},
\end{equation}
and $ F $ is the GGA enhancement factor.
Here, the modified Perdew-Burke-Ernzerhof GGA enhancement factor 
\begin{align}
  F \left( s \right)
  & =
    1 + \kappa
    -
    \frac{\kappa}{1 + \lambda \mu s^2 / \kappa}, \\
  \mu
  & = 
    0.21951,
    \qquad
    \kappa
    =
    0.804,
\end{align}
with $ \lambda = 1.25 $ \cite{Naito2019Phys.Rev.C99_024309} is used.
\subsection{Finite-size effects of nucleons}
\label{subsec:finite}
\par
In most works, protons and neutrons are assumed to be point particles,
and thus $ \rho_{\urm{ch}} \equiv \rho_p $ is assumed in the self-consistent steps
and the calculation of $ E_{\urm{gs}} $.
The difference between $ \rho_{\urm{ch}} $ and $ \rho_p $ is considered explicitly in the self-consistent steps in this paper.
Only the electric form factors of nucleons are considered,
while the magnetic form factors are not considered in the single-particle wave functions since
they appear in higher $ \left( 1/c^2 \right) $
and require us to consider the EM spin-orbit interaction self-consistently.
Instead, effects of the magnetic form factors are considered in the single-particle energy via the EM spin-orbit interaction, as discussed later.
\subsubsection{Electric form factor}
\par
The charge density distribution $ \rho_{\urm{ch}} $ is written
in terms of the electric form factors of protons and neutrons,
$ \tilde{G}_{\urm{E} p} $ and $ \tilde{G}_{\urm{E} n} $,
and the density distributions of protons and neutrons, $ \rho_p $ and $ \rho_n $ \cite{Ray1979Phys.Rev.C19_1855},
\begin{equation}
  \label{eq:char_den_mom}
  \tilde{\rho}_{\urm{ch}} \left( q \right)
  =
  \tilde{G}_{\urm{E} p} \left( q^2 \right) \,
  \tilde{\rho}_p \left( q \right)
  +
  \tilde{G}_{\urm{E} n} \left( q^2 \right) \,
  \tilde{\rho}_n \left( q \right),
\end{equation}
where the quantities with tilde denote those in the momentum space.
For example, $ \tilde{\rho}_p $ reads
\begin{widetext}
  \begin{equation}
    \label{eq:Frourier}
    \tilde{\rho}_p \left( q \right)
    =
    \frac{1}{\left( 2 \pi \right)^{3/2}}
    \int
    \rho_p \left( r \right) \,
    e^{-i \ve{q} \cdot \ve{r}}
    \, d \ve{r}
    =
    \sqrt{\frac{2}{\pi}}
    \int_0^{\infty}
    \rho_p \left( r \right) \,
    \frac{\sin \left( q r \right)}{qr}
    r^2 \, dr,
  \end{equation}
  with the Fourier transformation of $ \rho_p $.
  Here, the spherical symmetry is assumed for $ \rho_p $, $ \rho_n $, and $ \rho_{\urm{ch}} $.
  \par
  The electric form factors $ \tilde{G}_{\urm{E} p} $ and $ \tilde{G}_{\urm{E} n} $ are measured by the electron scattering of protons and neutrons,
  and their forms are taken from Ref.~\cite{Friedrich2003Eur.Phys.J.A17_607} as
  \begin{equation}
    \label{eq:Eform_mom}
    \tilde{G}_{\urm{E} \tau} \left( q^2 \right)
    =
    \frac{a_{10 \tau}}{\left( 1 + q^2 / a_{11 \tau} \right)^2}
    +
    \frac{a_{20 \tau}}{\left( 1 + q^2 / a_{21 \tau} \right)^2}
    +
    a_{b \tau} q^2
    \left[
      \exp \left\{
        -
        \frac{1}{2}
        \left(
          \frac{q - q_{b \tau}}{\sigma_{b \tau}}
        \right)^2
      \right\}
      +
      \exp \left\{
        -
        \frac{1}{2}
        \left(
          \frac{q + q_{b \tau}}{\sigma_{b \tau}}
        \right)^2
      \right\}
    \right],
  \end{equation}  
\end{widetext}
where the corresponding parameters are listed in Table \ref{tab:Eform}.
\par
In this paper, $ \rho_{\urm{ch}} $ given in Eq.~\eqref{eq:char_den_mom} is used in both the self-consistent steps and the calculation of $ E_{\urm{gs}} $, instead of the point-particle approximation.
Precisely, 
$ \rho_p $ and $ \rho_n $ are calculated from the single-particle wave functions $ \psi_i $,
and $ \rho_{\urm{ch}} $ is calculated from Eq.~\eqref{eq:char_den_mom} in each self-consistent step.
The effective potential $ V_{\urm{eff} \tau} $ for $ \psi_i $ is derived from $ \rho_{\urm{ch}} $ as well as $ \rho_p $ and $ \rho_n $.
As a result, 
the Coulomb potential in $ V_{\urm{eff} \tau} $ is different from that calculated in the point-particle approximation.
Details are shown as follows.
\begin{table*}
  \centering
  \caption{
    Parameters of the electric form factors $ \tilde{G}_{\urm{E} p} $ and $ \tilde{G}_{\urm{E} n} $
    taken from Ref.~\cite{Friedrich2003Eur.Phys.J.A17_607}.
    Uncertainty is not shown in this table.}
  \label{tab:Eform}
  \begin{ruledtabular}
    \begin{tabular}{lddddddd}
      \multicolumn{1}{c}{$ \tau $} & \multicolumn{1}{c}{$ a_{10 \tau} $} & \multicolumn{1}{c}{$ a_{11 \tau} $ ($ \mathrm{GeV}^2/c^2 $)} & \multicolumn{1}{c}{$ a_{20 \tau} $} & \multicolumn{1}{c}{$ a_{21 \tau} $ ($ \mathrm{GeV}^2/c^2 $)} & \multicolumn{1}{c}{$ a_{b \tau} $ ($ c^2 / \mathrm{GeV}^2 $)} & \multicolumn{1}{c}{$ q_{b \tau} $ ($ \mathrm{GeV} / c $)} & \multicolumn{1}{c}{$ \sigma_{b \tau} $ ($ \mathrm{GeV} / c $)} \\ \hline
      Proton & 1.041 & 0.765 & -0.041 & 6.2 & -0.23 & 0.07 & 0.27 \\ 
      Neutron & 1.04 & 1.73 & -1.04 & 1.54 & 0.23 & 0.29 & 0.20 
    \end{tabular}
  \end{ruledtabular}
\end{table*}
\subsubsection{Effective potential with finite-size effects}
\par
The effective potential of the nucleon $ \tau $ is, in general, defined as \cite{Vautherin1972Phys.Rev.C5_626, Engel2011_Springer-Verlag}
\begin{equation}
  \label{eq:eff_pot_def}
  V_{\urm{eff} \tau} \left( \ve{r} \right)
  =
  \frac{\delta E \left[\rho_p, \rho_n \right]}{\delta \rho_{\tau} \left( \ve{r} \right)}.
\end{equation}
Once the finite-size effects are considered, i.e.,~$ \rho_{\urm{ch}} \not \equiv \rho_p $,
the chain rule of the functional derivative \cite{Engel2011_Springer-Verlag}
\begin{equation}
  \label{eq:chain}
  \frac{\delta}{\delta f \left( \ve{r} \right)}
  =
  \int
  \frac{\delta g \left( \ve{r}' \right)}{\delta f \left( \ve{r} \right)}
  \frac{\delta}{\delta g \left( \ve{r}' \right)}
  \, d \ve{r}'
\end{equation}
should be applied to the Coulomb terms $ E_{\urm{Cd}} $ and $ E_{\urm{Cx}} $.
\par
The charge density distribution in the real space is
\begin{widetext}
  \begin{equation}
    \label{eq:char_den_real}
    \rho_{\urm{ch}} \left( r \right)
    =
    \frac{1}{\left( 2 \pi \right)^{3/2}}
    \left[
      \int 
      G_{\urm{E} p} \left( \left| \ve{r} - \ve{r}' \right| \right) \,
      \rho_p \left( r' \right)
      \, d \ve{r}'
      +
      \int 
      G_{\urm{E} n} \left( \left| \ve{r} - \ve{r}' \right| \right) \,
      \rho_n \left( r' \right)
      \, d \ve{r}' 
    \right],
  \end{equation}
\end{widetext}
where $ G_{\urm{E} \tau} $ are the electric form factors of the nucleons in the real space defined as
\begin{equation}
  \label{eq:Eform_real}
  G_{\urm{E} \tau} \left( r \right)
  = 
  \sqrt{\frac{2}{\pi}}
  \int_0^{\infty} 
  \tilde{G}_{\urm{E} \tau} \left( q^2 \right) \,
  \frac{\sin \left( q r \right)}{qr}
  q^2 \, dq,
\end{equation}
since the product in the momentum space is identical to the convolution in the real space as
\begin{align}
  \tilde{f} \left( q \right) \,
  \tilde{g} \left( q \right) 
  & =
    \sqrt{\frac{2}{\pi}}
    \int_0^{\infty}
    \left[
    \frac{1}{\left( 2 \pi \right)^{3/2}}
    \left( f * g \right)
    \left( r \right) 
    \right]
    \frac{\sin \left( q r \right)}{qr}
    r^2 \, dr, \\
  \left( 
  f * g
  \right)
  \left( r \right) 
  & =
    \int 
    f \left( \left| \ve{r} - \ve{r}' \right| \right) \,
    g \left( r' \right) \,
    d \ve{r}'.
\end{align}
\par
From Eqs.~\eqref{eq:chain} and \eqref{eq:char_den_real},
the functional derivative with respect to $ \rho_{\urm{ch}} $ reads
\begin{align}
  \frac{\delta}{\delta \rho_{\tau} \left( r \right)}
  & =
    \int
    \frac{\delta \rho_{\urm{ch}} \left( r' \right)}{\delta \rho_{\tau} \left( r \right)}
    \frac{\delta}{\delta \rho_{\urm{ch}} \left( r' \right)}
    \, d \ve{r}' \notag \\
  & = 
    \frac{1}{\left( 2 \pi \right)^{3/2}}
    \int
    G_{\urm{E} \tau} \left( \left| \ve{r} - \ve{r}' \right| \right) \,
    \frac{\delta}{\delta \rho_{\urm{ch}} \left( r' \right)}
    \, d \ve{r}'.
    \label{eq:chain_char}
\end{align}
Combining Eqs.~\eqref{eq:eff_pot_def} and \eqref{eq:chain_char},
the Coulomb potential for nucleons with the finite-size effects reads
\begin{align}
  V_{\urm{C} \tau} \left( r \right)
  & = 
    \frac{\delta E_{\urm{C}} \left[ \rho_{\urm{ch}} \right]}{\delta \rho_{\tau} \left( r \right)} \notag \\
  & =
    \int
    \frac{\delta E_{\urm{C}} \left[ \rho_{\urm{ch}} \right]}{\delta \rho_{\urm{ch}} \left( r' \right)} 
    \frac{\delta \rho_{\urm{ch}} \left( r' \right)}{\delta \rho_{\tau} \left( r \right)} 
    \, dr' \notag \\
  & =
    \frac{1}{\left( 2 \pi \right)^{3/2}}
    \int 
    \ca{V}_{\urm{C}} \left[ \rho_{\urm{ch}} \right]
    \left( r' \right) \,
    G_{\urm{E} \tau} \left( \left| \ve{r} - \ve{r}' \right| \right)
    \, d \ve{r}',
    \label{eq:pot_Coul_finite_real}
\end{align}
where 
$ E_{\urm{C}} \left[ \rho_{\urm{ch}} \right] = E_{\urm{Cd}} \left[ \rho_{\urm{ch}} \right] + E_{\urm{Cx}} \left[ \rho_{\urm{ch}} \right] $, 
and $ \ca{V}_{\urm{C}} $ is the conventional form of the Coulomb potential,
but expressing in terms of $ \rho_{\urm{ch}} $ instead, e.g.,
\begin{align}
  \ca{V}_{\urm{C}} \left[ \rho_{\urm{ch}} \right]
  \left( \ve{r} \right)
  & =
    \ca{V}_{\urm{Cd}} \left[ \rho_{\urm{ch}} \right]
    \left( \ve{r} \right)
    + 
    \ca{V}_{\urm{Cx}} \left[ \rho_{\urm{ch}} \right]
    \left( \ve{r} \right) \notag \\
  & =
    e^2
    \int
    \frac{\rho_{\urm{ch}} \left( \ve{r}' \right)}{\left| \ve{r} - \ve{r}' \right|}
    \, d \ve{r}'
    -
    e^2
    \left( \frac{3}{\pi} \right)^{1/3}
    \left[ \rho_{\urm{ch}} \left( \ve{r} \right) \right]^{1/3}
\end{align}
in the LDA form.
For the GGA form, see Ref.~\cite{Naito2019Phys.Rev.C99_024309}.
It should be noted that the Coulomb potential in the momentum space is
\begin{equation}
  \label{eq:pot_Coul_finite_mom}
  \tilde{V}_{\urm{C} \tau} \left[ \rho_{\urm{ch}} \right]
  \left( q \right)
  =
  \tilde{\ca{V}}_{\urm{C}} \left[ \rho_{\urm{ch}} \right]
  \left( q \right) \,
  \tilde{G}_{\urm{E} \tau} \left( q^2 \right).
\end{equation}
\par
Let us compare the Coulomb potential with the finite-size effects proposed in this paper
and that used in previous works such as Refs.~\cite{Chamel2009Phys.Rev.C80_065804, Auerbach1992Phys.Lett.B282_263, Roca-Maza2018Phys.Rev.Lett.120_202501}.
The Coulomb potential without the finite-size effects corresponds to $ \ca{V}_{\urm{C}} $ calculated with $ \rho_p $,
i.e., $ \ca{V}_{\urm{C}} \left[ \rho_p \right] $.
The Coulomb potential with the finite-size effects in previous works 
\cite{Chamel2009Phys.Rev.C80_065804, Auerbach1992Phys.Lett.B282_263, Roca-Maza2018Phys.Rev.Lett.120_202501}
is calculated with $ \rho_{\urm{ch}} $ when the potential is calculated,
i.e.,~$ \ca{V}_{\urm{C}} \left[ \rho_{\urm{ch}} \right] $.
This corresponds to $ \tilde{G}_{\urm{E} p} \equiv 1 $ and $ \tilde{G}_{\urm{E} n} \equiv 0 $ being used in Eq.~\eqref{eq:pot_Coul_finite_mom},
in which case the self-consistency shown in Eqs.~\eqref{eq:eff_pot_def} and \eqref{eq:chain_char} is no longer valid.
In contrast, the Coulomb potential with the finite-size effects in this work is derived self-consistently.
Hereafter, the finite-size effects in the previous works are simply called ``conventional finite-size effects,''
and those in this work are ``self-consistent finite-size effects.''
It should be emphasized that the Coulomb potential for the neutrons, $ V_{\urm{C} n} $, does not vanish within the self-consistent finite-size effects
since $ G_{\urm{E} n} \not \equiv 0 $,
whereas the conventional Coulomb potential $ \ca{V}_{\urm{C}} $ affects only protons.
\subsection{Vacuum polarization}
\par
The vacuum polarization is the lowest-order correction of quantum electrodynamics (QED) for the Coulomb interaction \cite{Weinberg1995_CambridgeUniversityPress}.
The effective one-body potential of the vacuum polarization for a charged particle $ -e $ under the Coulomb potential caused by the charge distribution $ \rho_{\urm{ch}} $ is known as the Uehling potential \cite{Uehling1935Phys.Rev.48_55}.
In the case of atomic nuclei, the charge of protons is $ +e $,
and thus the sign of the potential is opposite to the original Uehling potential as
\begin{equation}
  \label{eq:Uehling_orig}
  V_{\urm{VP}} \left( \ve{r} \right)
  =
  \frac{2}{3}
  \frac{\alpha e^2}{\pi}
  \int
  \frac{\rho_{\urm{ch}} \left( \ve{r}' \right)}{\left| \ve{r} - \ve{r}' \right|}
  \ca{K}_1
  \left(
    \frac{2}{\lambdabar_e}
    \left| \ve{r} - \ve{r}' \right|
  \right)
  \, d \ve{r}',
\end{equation}
where 
\begin{equation}
  \ca{K}_1
  \left( x \right)
  =
  \int_1^{\infty}
  e^{-xt}
  \left(
    \frac{1}{t^2}
    +
    \frac{1}{2t^4}
  \right)
  \sqrt{t^2 - 1}
  \, dt,
\end{equation}
$ \alpha $ is the fine-structure constant,
and 
$ \lambdabar_e = 386.15926796 \, \mathrm{fm} $ is the reduced Compton wavelength of electrons
\footnote{
  The dominant contribution of a virtual particle-antiparticle pair produced in a photon propagator is the lightest fermions, i.e.,~electrons.
  Therefore, quantities appeared in the Uehling potential are still those for electrons
  even though the potential is applied to protons.} \cite{CODATA}.
Correspondingly, the EDF for the vacuum polarization is written as
\begin{equation}
  E_{\urm{VP}} \left[ \rho_{\urm{ch}} \right]
  =
  \frac{1}{2}
  \int 
  \rho_{\urm{ch}} \left( \ve{r} \right) \, 
  V_{\urm{VP}} \left( \ve{r} \right) \, 
  d \ve{r}.
\end{equation}
\par
Once spherical symmetry is assumed,
Eq.~\eqref{eq:Uehling_orig} is written as \cite{WayneFullerton1976Phys.Rev.A13_1283}
\begin{widetext}
  \begin{equation}
    \label{eq:Uehling_sphe}
    V_{\urm{VP}} \left( r \right)
    =
    \frac{2 \alpha e^2 \lambdabar_e}{3r}
    \int_0^{\infty}
    \left[
      \ca{K}_0
      \left(
        \frac{2}{\lambdabar_e}
        \left| r - r' \right|
      \right)
      -
      \ca{K}_0
      \left(
        \frac{2}{\lambdabar_e}
        \left| r + r' \right|
      \right)
    \right]
    \rho_{\urm{ch}} \left( r' \right) \,
    r' \, dr',
  \end{equation}
\end{widetext}
where
\begin{align}
  \ca{K}_0 \left( x \right)
  & =
    - \int_{- \infty}^x
    \ca{K}_1 \left( x' \right) \, dx' \notag \\
  & =
    \int_1^{\infty}
    e^{-xt}
    \left(
    \frac{1}{t^3}
    +
    \frac{1}{2t^5}
    \right)
    \sqrt{t^2 - 1}
    \, dt.
\end{align}
\par
In this work, this potential is assumed to affect only protons.
According to the treatment of the finite-size effects in the DFT scheme discussed in Sec.~\ref{subsec:finite},
the vacuum polarization potential for protons and neutrons, $ V_{\urm{VP} \tau} $, are $ \delta E_{\urm{VP}} / \delta \rho_{\tau} $,
slightly different from the original $ V_{\urm{VP}} $.
However, this difference must be tiny.
Therefore, $ V_{\urm{VP}} $ is applied to protons, and the vacuum polarization potential for neutrons is neglected,
while $ \rho_{\urm{ch}} $ is used for calculating $ V_{\urm{VP}} $ in Eq.~\eqref{eq:Uehling_sphe}.
This corresponds to the conventional treatment of the finite-size effects in the Coulomb potential.
\subsection{Electromagnetic spin-orbit interaction}
\par
The protons and neutrons move inside the charge distribution $ \rho_{\urm{ch}} $.
On the frame of a nucleon, this charge distribution is regarded as moving
and the moving charge distribution generates a magnetic field,
which interacts with the spin of the nucleon.
This interaction is the spin-orbit interaction.
\par
In this work, the EM spin-orbit interaction is considered by using the first-order perturbation theory.
The effects on the single-particle orbitals and potentials are neglected
since this interaction affects the single-particle energies by less than $ 100 \, \mathrm{keV} $.
The correction due to the EM spin-orbit interaction
for the single-particle energy is \cite{Roca-Maza2018EPJWebConf.194_01002}
\begin{widetext}
  \begin{align}
    \Delta \epsilon_i
    & =
      \frac{\hbar^2 c^2}{2 m^2 c^4}
      x_i
      \avr{\hat{\ve{l}}_i \cdot \hat{\ve{s}}_i}
      \int_0^{\infty}
      \frac{\left[ u_i \left( r \right) \right]^2}{r}
      \frac{d \ca{V}_{\urm{C}} \left( r \right)}{dr}
      \, dr \notag \\
    & =
      \frac{\hbar^2 c^2}{2 m^2 c^4}
      x_i
      \left[
      j_i \left(j_i + 1 \right)
      -
      l_i \left(l_i + 1 \right)
      -
      \frac{3}{4}
      \right]
      \int_0^{\infty}
      \frac{\left[ u_i \left( r \right) \right]^2}{r}
      \frac{d \ca{V}_{\urm{C}} \left( r \right)}{dr}
      \, dr ,
      \label{eq:SP}
  \end{align}
\end{widetext}
where
$ \hat{\ve{l}}_i $ and $ \hat{\ve{s}}_i $ are its orbital and spin angular-momentum operators,
$ r u_i \left( r \right) $ is the radial part of single-particle wave function, 
and $ l_i $ and $ j_i $ are the azimuthal quantum number and total angular momentum, respectively.
The quantity $ x_i $ is related to the $ g $ factors as \cite{CODATA}
\begin{equation}
  \label{eq:g-factor}
  x_i
  =
  \begin{cases}
    g_p - 1 = 4.5856946893 & \text{for protons}, \\
    g_n     = -3.82608545 & \text{for neutrons}.
  \end{cases}
\end{equation}
This equation is the same as the spin-orbit interaction for hydrogen-like atoms \cite{Greiner1998QuantumMechanics---SpecialChapters_Springer-Verlag},
while the $ g $ factors of nucleons are used instead.
Here, $ 1 $ in Eq.~\eqref{eq:g-factor} corresponds to the charge of protons.
In this calculation, the Coulomb potential without the finite-size effects $ \ca{V}_{\urm{C}} \left[ \rho_p \right] $ is used,
since the correction itself is expected to be small, and thus it does not need to take into account the finite-size correction in $ \ca{V}_{\urm{C}} \left[ \rho_{\urm{ch}} \right] $.
%
\section{Simple Estimation of Systematic Behaviors}
\label{sec:simple}
\par
Before the numerical calculation, 
simple estimations are performed in this section
to understand the systematic behavior of the contributions of
the finite-size effects and the vacuum polarization to the total energy.
The hard-sphere distribution is assumed for protons:
\begin{equation}
  \rho_p \left( r \right)
  =
  \begin{cases}
    \rho_0^p
    & r < R_p, \\
    0
    & r > R_p,
  \end{cases}
\end{equation}
where $ R_p $ is the radius of the proton distribution
and
\begin{equation}
  \rho_0^p
  =
  \frac{3Z}{4 \pi R_p^3}
\end{equation}
is held.
The saturation density of protons, 
\begin{equation}
  \label{eq:hard_rhop}
  \rho_0^p
  =
  \frac{1}{2} \rho_0
  \simeq
  0.08 \, \mathrm{fm}^{-3} ,
\end{equation}
together with
\begin{equation}
  \label{eq:Rp}
  R_p
  =
  \left(
    \frac{3Z}{4 \pi \rho_0^p}
  \right)^{1/3},
\end{equation}
are
used for estimation of coefficients,
where $ \rho_0 $ is the saturation density of atomic nuclei.
Accordingly, the neutron and charge distributions are also assumed to be hard spheres.
When the finite-size effects are considered,
we use proton and neutron radii, $ \avr{r_p^2} $ and $ \avr{r_n^2} $, consistent with the form factors \eqref{eq:Eform_real}.
Note that smaller or larger values of proton radius
that are recently debated in the literature do not affect our simple estimation significantly.
%
\par
At first, the estimation under the point-particle approximation is discussed.
The Coulomb direct potential is
\begin{align}
  V_{\urm{Cd}}^{\urm{point}} \left( r \right)
  & =
    e^2
    \int
    \frac{\rho_p \left( r' \right)}{\left| \ve{r} - \ve{r}' \right|}
    \, d \ve{r}' \notag \\
  & =
    \begin{cases}
      \frac{Ze^2}{2 R_p} \left( 3 - \frac{r^2}{R_p^2} \right)
      & r < R_p, \\
      \frac{Ze^2}{r}
      & r > R_p,
    \end{cases}
\end{align}
and thus the Coulomb direct energy is
\begin{align}
  E_{\urm{Cd}}^{\urm{point}}
  & =
    \frac{1}{2}
    \int
    \rho_p \left( r \right) \,
    V_{\urm{Cd}}^{\urm{point}} \left( r \right) \,
    d \ve{r} \notag \\
  & =
    \frac{3e^2}{5}
    \frac{Z^2}{R_p} \notag \\
  & =
    \frac{3e^2}{5}
    \left( \frac{4 \pi \rho_0^p}{3} \right)^{1/3}
    Z^{5/3}
    \notag \\
  & \simeq
    0.60
    Z^{5/3}
    \, \mathrm{MeV}.
    \label{eq:estimate_ECd}
\end{align}
The Coulomb exchange energy is 
\begin{align}
  E_{\urm{Cx}}^{\urm{point}}
  & =
    -
    \frac{3 e^2}{4}
    \left( \frac{3}{\pi} \right)^{1/3}
    \int
    \left[
    \rho_p \left( r \right)
    \right]^{4/3}
    \, d \ve{r} \notag \\
  & =
    -
    \frac{3 e^2}{4}
    \left( \frac{9}{4 \pi^2} \right)^{1/3}
    \frac{Z^{4/3}}{R_p} \notag \\
  & =
    -
    \frac{3 e^2}{4}
    \left( \frac{3 \rho_0^p}{\pi} \right)^{1/3}
    Z
    \notag \\
  & \simeq
    -0.46 Z
    \, \mathrm{MeV}.
    \label{eq:estimate_ECx}
\end{align}
\par
Let us consider the finite-size effects for the Coulomb energy.
Hereafter, the superscripts associated with energies describe which finite-size effects are considered;
``point,'' ``$ p $-finite,'' and ``$ pn $-finite'' mean the energies calculated with the point-particle approximation, the proton finite-size effect, and both the proton and the neutron finite-size effects, respectively.
Only the finite-size correction to the Coulomb direct energy is discussed here,
since the finite-size effects should be a small correction 
and 
the main contribution of the Coulomb energy to the total energy is the direct term.
The relationship between the radii is assumed to be \cite{Chabanat1997Nucl.Phys.A627_710}
\begin{equation}
  \label{eq:radii}
  R_{\urm{ch}}^2
  \simeq
  R_p^2
  +
  \avr{r_p^2}
  +
  \frac{N}{Z}
  \avr{r_n^2},
\end{equation}
where $ R_{\urm{ch}} $ is the charge radius of the nucleus.
Here, the contribution of the EM spin-orbit interaction is not considered in Eq.~\eqref{eq:radii}.
The Coulomb direct energy with the finite-size effects is the same as Eq.~\eqref{eq:estimate_ECd},
while $ R_{\urm{ch}} $ is used instead of $ R_p $, i.e.,
\begin{equation}
  \label{eq:ECd_finite}
  E_{\urm{Cd}}^{\urm{finite}}
  = 
  \frac{3e^2}{5}
  \frac{Z^2}{R_{\urm{ch}}} .
\end{equation}
The contribution of the proton finite-size effect for the total energy is estimated
with $ \avr{r_n^2} = 0 $ in $ R_{\urm{ch}} $ of Eq.~\eqref{eq:radii} as
\begin{align}
  E_{\urm{C}}^{\urm{$ p $-finite}}
  -
  E_{\urm{C}}^{\urm{point}}
  & \simeq
    E_{\urm{Cd}}^{\urm{$ p $-finite}}
    -
    E_{\urm{Cd}}^{\urm{point}}
    \notag \\
  & = 
    \frac{3e^2}{5}
    Z^2
    \left(
    \frac{1}{R_{\urm{ch}}^{\urm{$ p $-finite}}}
    -
    \frac{1}{R_p}
    \right)
    \notag \\
  & = 
    \frac{3e^2}{5}
    Z^2
    \left[
    \frac{1}{
    \sqrt{R_p^2 + \avr{r_p^2}}}
    -
    \frac{1}{R_p}
    \right]
    \notag \\
  & \simeq
    -
    \frac{3e^2 \avr{r_p^2}}{10}
    \frac{Z^2}{R_p^3}
    \notag \\
  & =
    -
    \frac{2 \pi e^2 \rho_0^p \avr{r_p^2}}{5}
    Z
    \notag \\
  & \simeq
    -0.11 Z
    \, \mathrm{MeV}.
    \label{eq:estimate_Ep}
\end{align}
The contribution of the neutron finite-size effect to the total energy is 
\begin{align}
  & E_{\urm{C}}^{\urm{$ pn $-finite}}
    -
    E_{\urm{C}}^{\urm{$ p $-finite}}
    \notag \\
  & \simeq
    E_{\urm{Cd}}^{\urm{$ pn $-finite}}
    -
    E_{\urm{Cd}}^{\urm{$ p $-finite}}
    \notag \\
  & = 
    \frac{3e^2}{5}
    Z^2
    \left(
    \frac{1}{R_{\urm{ch}}^{\urm{$ pn $-finite}}}
    -
    \frac{1}{R_{\urm{ch}}^{\urm{$ p $-finite}}}
    \right)
    \notag \\
  & = 
    \frac{3e^2}{5}
    Z^2
    \left[
    \frac{1}{
    \sqrt{R_p^2 + \avr{r_p^2} + \frac{N}{Z} \avr{r_n^2}}}
    -
    \frac{1}{
    \sqrt{R_p^2 + \avr{r_p^2}}}
    \right]
    \notag \\
  & \simeq
    -
    \frac{3e^2 \avr{r_n^2}}{10}
    \frac{NZ}{\left( R_p^2 + \avr{r_p^2} \right)^{3/2}}
    \notag \\
  & \simeq
    -
    \frac{3e^2 \avr{r_n^2}}{10}
    \frac{NZ}{R_p^3}
    \notag \\
  & =
    -
    \frac{2 \pi e^2 \rho_0^p \avr{r_n^2}}{5}
    N
    \notag \\
  & \simeq
    0.010 N
    \, \mathrm{MeV}.
    \label{eq:estimate_En}
\end{align}
Since $ \avr{r_p^2} > 0 $ and $ \avr{r_n^2} < 0 $, 
the coefficient in Eq.~\eqref{eq:estimate_En} is positive,
whereas that in Eq.~\eqref{eq:estimate_Ep} is negative.
\par
At the end of this section, the energy of the vacuum polarization is estimated.
It reads
\begin{widetext}
  \begin{align}
    E_{\urm{VP}}
    & =
      \frac{1}{2}
      \int
      \rho_{\urm{ch}} \left( r \right) \, 
      V_{\urm{VP}} \left( r \right) \,
      d \ve{r}
      \notag \\
    & \simeq
      2 \pi
      \int_0^{\infty}
      \rho_p \left( r \right) \, 
      V_{\urm{VP}} \left( r \right) \,
      r^2 \, dr
      \notag \\
    & = 
      2 \pi \rho_0^p
      \int_0^{R_p}
      V_{\urm{VP}} \left( r \right) \,
      r^2 \, dr
      \notag \\
    & =
      2 \pi \rho_0^p
      \frac{2 \alpha e^2 \lambdabar_e}{3}
      \int_0^{R_p}
      \int_0^{\infty}
      \left[
      \ca{K}_0
      \left(
      \frac{2}{\lambdabar_e}
      \left| r - r' \right|
      \right)
      -
      \ca{K}_0
      \left(
      \frac{2}{\lambdabar_e}
      \left| r + r' \right|
      \right)
      \right]
      \rho_{\urm{ch}} \left( r' \right) \,
      r' \, dr'
      \, r \, dr
      \notag \\
    & \simeq
      2 \pi \left( \rho_0^p \right)^2
      \frac{2 \alpha e^2 \lambdabar_e}{3}
      \int_0^{R_p}
      \int_0^{R_p}
      \left[
      \ca{K}_0
      \left(
      \frac{2}{\lambdabar_e}
      \left| r - r' \right|
      \right)
      -
      \ca{K}_0
      \left(
      \frac{2}{\lambdabar_e}
      \left| r + r' \right|
      \right)
      \right]
      r' \, dr'
      \, r \, dr.
      \label{eq:estimate_VP_1}
  \end{align}
  The first term of the integral in Eq.~\eqref{eq:estimate_VP_1} is estimated as 
  \begin{align}
    & \int_0^{R_p}
      \int_0^{R_p}
      \ca{K}_0
      \left(
      \frac{2}{\lambdabar_e}
      \left| r - r' \right|
      \right)
      r' \, dr'
      \, r \, dr
      \notag \\
    & =
      \int_0^{R_p}
      \int_0^{R_p}
      \int_1^{\infty}
      e^{-2c \left| r - r' \right|}
      \left(
      \frac{1}{t^3}
      +
      \frac{1}{2t^5}
      \right)
      \sqrt{t^2 - 1}
      \, dt 
      \, r' \, dr'
      \, r \, dr
      \notag \\
    & =
      \int_1^{\infty}
      \int_0^{R_p}
      \left[
      \int_0^r
      e^{-2c \left( r - r' \right)}
      r' \, dr'
      +
      \int_r^{R_p}
      e^{2c \left( r - r' \right)}
      r' \, dr'
      \right]
      r \, dr
      \, \left(
      \frac{1}{t^3}
      +
      \frac{1}{2t^5}
      \right)
      \sqrt{t^2 - 1}
      \, dt 
      \notag \\
    & =
      \int_1^{\infty}
      \int_0^{R_p}
      \frac{r}{4c^2}
      \left[
      4cr 
      +
      e^{-2cr}
      -
      e^{2c \left( r - R_p \right)}
      \left( 1 + 2cR_p \right)
      \right]
      \, dr
      \, \left(
      \frac{1}{t^3}
      +
      \frac{1}{2t^5}
      \right)
      \sqrt{t^2 - 1}
      \, dt 
      \notag \\
    & =
      \int_1^{\infty}
      \frac{3 - 6c^2 R_p^2 + 8c^3 R_p^3 - 3e^{-2cR_p} \left( 1 + 2cR_p \right)}{24c^4}
      \left(
      \frac{1}{t^3}
      +
      \frac{1}{2t^5}
      \right)
      \sqrt{t^2 - 1}
      \, dt ,
      \label{eq:estimate_VP_2}
  \end{align}
  and the second term is
  \begin{align}
    \int_0^{R_p}
    \int_0^{R_p}
    \ca{K}_0
    \left(
    \frac{2}{\lambdabar_e}
    \left| r + r' \right|
    \right)
    r' \, dr'
    \, r \, dr
    & =
      \int_0^{R_p}
      \int_0^{R_p}
      \int_1^{\infty}
      e^{-2c \left| r + r' \right|}
      \left(
      \frac{1}{t^3}
      +
      \frac{1}{2t^5}
      \right)
      \sqrt{t^2 - 1}
      \, dt 
      \, r' \, dr'
      \, r \, dr
      \notag \\
    & =
      \int_1^{\infty}
      \int_0^{R_p}
      \int_0^{R_p}
      e^{-2c \left( r + r' \right)}
      r' \, dr'
      \, r \, dr
      \, \left(
      \frac{1}{t^3}
      +
      \frac{1}{2t^5}
      \right)
      \sqrt{t^2 - 1}
      \, dt 
      \notag \\
    & =
      \int_1^{\infty}
      \frac{e^{-4cR_p} \left(1 - e^{2cR_p} + 2cR_p \right)^2}{16c^4}
      \left(
      \frac{1}{t^3}
      +
      \frac{1}{2t^5}
      \right)
      \sqrt{t^2 - 1}
      \, dt ,
      \label{eq:estimate_VP_3}
  \end{align}
  where $ c = t / \lambdabar_e $.
  Combining Eqs.~\eqref{eq:estimate_VP_1}--\eqref{eq:estimate_VP_3},
  we get 
  \begin{align}
    & \int_0^{R_p}
      \int_0^{R_p}
      \left[
      \ca{K}_0
      \left(
      \frac{2}{\lambdabar_e}
      \left| r - r' \right|
      \right)
      -
      \ca{K}_0
      \left(
      \frac{2}{\lambdabar_e}
      \left| r + r' \right|
      \right)
      \right]
      r' \, dr'
      \, r \, dr
      \notag \\
    & =
      \int_1^{\infty}
      \left[
      \frac{3 - 6c^2 R_p^2 + 8c^3 R_p^3 - 3e^{-2cR_p} \left( 1 + 2cR_p \right)}{24c^4}
      -
      \frac{e^{-4cR_p} \left(1 - e^{2cR_p} + 2cR_p \right)^2}{16c^4}
      \right]
      \left(
      \frac{1}{t^3}
      +
      \frac{1}{2t^5}
      \right)
      \sqrt{t^2 - 1}
      \, dt
      \notag \\
    & \simeq
      0.0070 R_p^5 \, \mathrm{MeV}.
  \end{align}
\end{widetext}
Finally, we can estimate Eq.~\eqref{eq:estimate_VP_1} as 
\begin{align}
  E_{\urm{VP}}
  & \simeq
    2 \pi \left( \rho_0^p \right)^2
    \frac{2 \alpha e^2 \lambdabar_e}{3}
    \times
    0.0070 R_p^5 \, \mathrm{MeV}
    \notag \\
  & \simeq
    0.0047 Z^{5/3} \, \mathrm{MeV}.
    \label{eq:estimate_VP}
\end{align}
\par
At the end of this section, effects of these terms on the Coulomb displacement energy are discussed.
The displacement energy is defined as \cite{Auerbach1972Rev.Mod.Phys.44_48}
\begin{equation}
  E_{\urm{dis}}
  =
  \frac{\brakket{\mathrm{P}}{T_{+} \left[ H, T_{-} \right]}{\mathrm{P}}}{N},
\end{equation}
where $ \ket{\mathrm{P}} $ is the parent nucleus,
$ H $ is the total Hamiltonian, and
$ T_{\pm} $ are the isospin raising and lowering operators.
Auerbach \textit{et al\/.}~\cite{Auerbach1972Rev.Mod.Phys.44_48}
estimated the contributions of
the Coulomb exchange, proton finite-size effect, and the vacuum polarization as
$ -900 Z/A \, \mathrm{keV} $, $ \sim - 100 \, \mathrm{keV} $,
and $ 8.5 Z / A^{1/3} \, \mathrm{keV} $, respectively.
\par
Each contribution of the Coulomb energy to the displacement energy is calculated from the simple estimations as
$ E_{\urm{C}i}^{\urm{D}} - E_{\urm{C}i}^{\urm{P}} $,
where $ E_{\urm{C}i}^{\urm{P}} $ and $ E_{\urm{C}i}^{\urm{D}} $ refer to the energies of each contribution for the parent and daughter nuclei, respectively.
Since the proton and neutron numbers $ Z $ and $ N $ change to $ Z + 1 $ and $ N - 1 $ in the isobaric analog resonance,
the contributions to the displacement energy of
the Coulomb exchange, proton finite-size effect, neutron finite-size effect, and the vacuum polarization read approximately
$ -460 \, \mathrm{keV} $, $ -110 \, \mathrm{keV} $, $ -10 \, \mathrm{keV} $, and $ 7.8 Z^{2/3} \, \mathrm{keV} $, respectively.
These values are consistent with Auerbach's estimation if $ Z \simeq A/2 $ is assumed.
\section{Theoretical Calculation and Discussion}
\label{sec:calc}
\par
The finite-size effects, the vacuum polarization, and the EM spin-orbit interaction
are implemented to the self-consistent Skyrme Hartree-Fock plus RPA code named
\textsc{skyrme\_rpa} \cite{Colo2013Comput.Phys.Commun.184_142}.
The ISB terms of the nuclear force are also implemented 
to compare the contributions of the EM interaction with that of the ISB terms of nuclear force.
Details of the ISB terms of nuclear force are shown in the appendix.
In this calculation, spherical symmetry is assumed, 
and the pairing correlations are not considered.
A box of $ 15 \, \mathrm{fm} $ with $ 0.1 \, \mathrm{fm} $ mesh is used.
In each self-consistent step,
the charge density distribution and the Coulomb potential are calculated with Eqs.~\eqref{eq:char_den_mom} and \eqref{eq:pot_Coul_finite_mom}, respectively,
instead of the convolution in the real space.
More precisely, in each self-consistent step, the nucleon densities $ \rho_{\tau} $ and the Coulomb potential $ \ca{V}_{\urm{C}} $,
obtained by the \textsc{skyrme\_rpa} code in the real space,
are transformed to the momentum space,
and then $ \tilde{\rho}_{\urm{ch}} $ and $ \tilde{V}_{\urm{C}} $ are derived and transformed back to the real space.
\par
In this paper, the SAMi functional \cite{Roca-Maza2012Phys.Rev.C86_031306} is used for the nuclear EDF for most calculations,
and the SAMi-ISB functional \cite{Roca-Maza2018Phys.Rev.Lett.120_202501} is used instead when the isospin symmetry breaking originated from the nuclear interaction is considered explicitly.
\par
One may wonder whether the Skyrme functional should be refitted.
The coefficients of the Skyrme functionals are determined to reproduce the experimental total binding energies and density distributions of the selected nuclei.
In this paper, our main motivation is to see how these corrections affect the nuclear properties in the Skyrme Hartree-Fock calculations instead of the comparison between the calculations and experimental data.
In short, this is a sensitivity study.
If we wished to compare with experimental data in detail,
we would need a refit of the Skyrme functional.
\subsection{Systematic calculation}
\par
First of all, the systematic behavior of the contributions of
the finite-size effects, vacuum polarization, and EM spin-orbit interaction are discussed.
Some doubly magic and semimagic nuclei are selected.
The calculations are performed under the assumption of spherical symmetry without pairing, 
which is assumed not to affect the main conclusions of the present paper, even for the semi-magic nuclei.
\par
\begin{figure}[tb]
  \centering
  \includegraphics[width=1.0\linewidth]{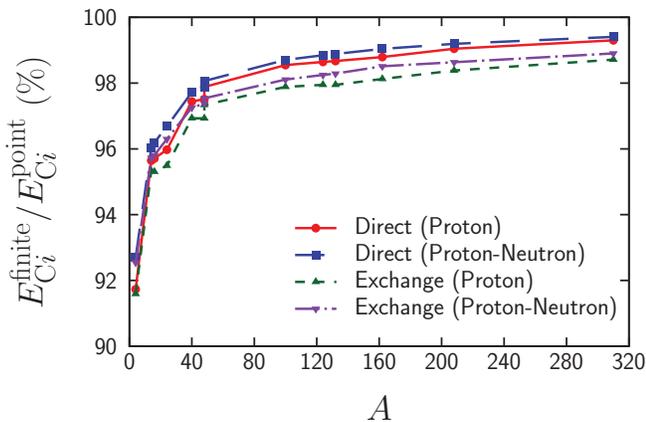}
  \caption{
    Ratios of the Coulomb direct and exchange energies calculated with finite-size effects to those without the finite-size effects,
    $ E_{\urm{C}i}^{\urm{finite}} / E_{\urm{C}i}^{\urm{point}} $
    ($ i = \mathrm{d} , \, \mathrm{x} $), shown as functions of the mass number $ A $.
    The ratios for the Coulomb direct and exchange terms only with the proton finite-size effect are shown by the red solid and green dashed lines, respectively.
    Those with both the proton and neutron finite-size effects are shown by the blue long-dashed and purple dot-dashed lines, respectively.}
  \label{fig:sys_deltaE}
\end{figure}

In Table \ref{tab:sys_tot}, the total energies of the selected nuclei are shown,
where contributions are considered step by step to see the effects of each term.
In Fig.~\ref{fig:sys_deltaE}, the ratios of the Coulomb direct and exchange energies calculated with the finite-size effects to those without the finite-size effects,
$ E_{\urm{C}i}^{\urm{finite}} / E_{\urm{C}i}^{\urm{point}} $
($ i = \mathrm{d} , \, \mathrm{x} $), are shown as functions of the mass number $ A $.
\par
It is seen that the proton finite-size effect makes the nuclei more bound,
for example, 
$ 580 \, \mathrm{keV} $ and $ 8.2 \, \mathrm{MeV} $ for $ \nuc{O}{16}{} $ and $ \nuc{Pb}{208}{} $, respectively.
In contrast, the neutron finite-size effect makes the nuclei less bound,
for instance, 
$ 64 \, \mathrm{keV} $ and $ 1.2 \, \mathrm{MeV} $ for $ \nuc{O}{16}{} $ and $ \nuc{Pb}{208}{} $, respectively.
\par
From the point of view of the interaction, the finite-size effect of protons makes the Coulomb interaction effectively weaker,
because $ \rho_{\urm{ch}} $ distributes more extensive and dilute than $ \rho_p $ due to the proton finite size as shown in Eq.~\eqref{eq:radii}.
In contrast, 
the finite-size effect of neutrons makes the Coulomb interaction effectively stronger, 
because the neutron radius $ \avr{r_n^2} $ is negative and hence
neutrons effectively behave as negative charge rather than as positive charge,
which make $ \rho_{\urm{ch}} $ distributes more compact and denser.
The absolute value of 
the proton radius $ \left| \avr{r_p^2} \right| $ is larger than
that of the neutron radius $ \left| \avr{r_n^2} \right| $,
and therefore, the proton finite-size effect is more significant than the neutron finite-size effect as shown in Fig.~\ref{fig:sys_deltaE}.
Although the finite-size effects for light nuclei are more significant
than those for the heavy nuclei since $ R_p $ is smaller, 
the absolute values themselves for the heavy nuclei are more significant.
It should be noted that even though the neutron finite-size effect is smaller than the proton one,
it is not small enough to be neglected in heavy nuclei.
\par
As shown in Table \ref{tab:sys_tot},
the vacuum polarization makes the nuclei less bound,
for example, 
$ 88 \, \mathrm{keV} $ and $ 3.7 \, \mathrm{MeV} $ for $ \nuc{O}{16}{} $ and $ \nuc{Pb}{208}{} $, respectively.
The vacuum polarization contributes to the total energy more than
the difference between the exact-Fock and the LDA Coulomb exchange energies
and the neutron finite-size effect.
Hence, it is not small at all to be neglected in heavy nuclei.
\par
Among the contributions of the finite-size effects and the vacuum polarization to the total energy,
the proton finite-size effect is the dominant.
The contribution of the vacuum polarization is larger than that of the neutron finite-size effect in the heavy nuclei,
whereas they are comparable in the light nuclei.
\par
Next, the systematic behaviors of these contributions to the total energy are discussed.
For this estimation, 
the Coulomb direct and exchange energies calculated with the point-nucleon approximation,
i.e.,~$ E_{\urm{Cd}}^{\urm{point}} $ and $ E_{\urm{Cx}}^{\urm{point}} $, are used.
The proton finite-size effect and vacuum polarization are defined as the differences of the two total energies,
i.e.,~$ E_{\urm{tot}}^{\urm{$ p $-finite}} - E_{\urm{tot}}^{\urm{point}} $
and $ E_{\urm{tot}}^{\urm{vacuum}} - E_{\urm{tot}}^{\urm{$ pn $-finite}} $
where $ E_{\urm{tot}}^{\urm{vacuum}} $ is the total energy calculated with the all finite-size effects and vacuum polarization but without the EM spin-orbit interaction.
All the energies used here are calculated with the LDA Coulomb functional.
These energies are fit to 
\begin{equation}
  \label{eq:sys_Z}
  E
  =
  a Z^b ,
\end{equation}
where as the neutron finite-size effect
defined in term of $ E_{\urm{tot}}^{\urm{$ pn $-finite}} - E_{\urm{tot}}^{\urm{$ p $-finite}} $
is fit to 
\begin{equation}
  \label{eq:sys_N}
  E
  =
  a N^b.
\end{equation}
These coefficients $ a $ and $ b $ are shown in Table \ref{tab:sys_para}.
\par
At first, the values of $ a $ and $ b $ are almost compatible with
the simple estimation, performed in Sec.~\ref{sec:simple}.
Also, as discussed above, 
the proton finite-size effect is one order of magnitude smaller than the exchange energy,
and the neutron finite-size effect and the vacuum polarization are one more order of magnitude smaller, according to
the values of $ a $ in Table \ref{tab:sys_para}.
Since the value of $ b $ for the vacuum polarization is larger than that for the neutron finite-size effect,
the contribution of the vacuum polarization to the total energy is larger than that of the neutron finite-size effect in the heavy nuclei, as discussed above.
\par
At last, the EM spin-orbit interaction gives different effects for the total energy in different nuclei.
As expected, in the spin-saturated nuclei, such as $ \nuc{He}{4}{} $, $ \nuc{O}{16}{} $, and $ \nuc{Ca}{40}{} $,
the EM spin-orbit interaction contributes to the total energy by only a few $ \mathrm{keV} $.
In contrast, in the spin-unsaturated nuclei, the absolute values of its contribution to the total energy are around $ 50 \, \mathrm{keV} $ or more.
For example, on the one hand, in $ \nuc{Ca}{48}{} $ case, 
$ \nu 1f_{7/2} $ orbital for neutrons is fully occupied,
while its spin-orbit partner $ \nu 1f_{5/2} $ orbital is completely unoccupied.
The coefficient of Eq.~\eqref{eq:SP} for the $ \nu 1f_{7/2} $ orbital is negative.
On the other hand, in the $ \nuc{Ni}{48}{} $ case, the $ \pi 1f_{7/2} $ orbital for protons is fully occupied,
while its spin-orbit partner $ \pi 1f_{5/2} $ orbital is completely unoccupied.
The coefficient of Eq.~\eqref{eq:SP} for $ \pi 1f_{7/2} $ orbital is positive.
Therefore, the contribution of the EM spin-orbit interaction to the total energies for $ \nuc{Ca}{48}{} $ and $ \nuc{Ni}{48}{} $ should be opposite,
because the corresponding contribution to the $ \nuc{Ca}{40}{} $ core is almost zero.
Indeed, as shown in Table \ref{tab:sys_tot}, its contributions for $ \nuc{Ca}{48}{} $ and $ \nuc{Ni}{48}{} $ are $ 139 \, \mathrm{keV} $ and $ -206 \, \mathrm{keV} $, respectively.
The absolute value itself reflects the structure of the single-particle wave functions,
in contrast with the other contributions which have just monotonic $ Z $ or $ N $ dependence.%
\begin{table*}[tb]
  \centering
  \caption{
    Total energies for selected doubly magic and semimagic nuclei.
    All the corrections to the Coulomb interaction are considered step by step,
    where the SAMi functional \cite{Roca-Maza2012Phys.Rev.C86_031306} is used for the nuclear EDF.
    The columns 
    labeled ``LDA'' and ``GGA'' refer to the results without the finite-size effects,
    while those labeled as ``$ \text{GGA} + \text{$ p $-fin} $'' and ``$ \text{GGA} + \text{$ pn $-fin} $'' refer to the results with proton and proton-neutron finite-size effects.
    Moreover, ``VP'' refers the vacuum polarization,
    and ``All'' corresponds to ``$\text{GGA} + \text{$ pn $-fin} + \text{VP} + \text{EM spin-orbit} $''.
    The column labeled as ``$ \text{All} + \text{ISB} $'' is calculated with all the corrections and the SAMi-ISB functional \cite{Roca-Maza2018Phys.Rev.Lett.120_202501}.
    All units are in $ \mathrm{MeV} $.}
  \label{tab:sys_tot}
  \begin{ruledtabular}
    \begin{tabular}{rddddddd}
      \multicolumn{1}{c}{Nuclei} & \multicolumn{1}{c}{LDA} & \multicolumn{1}{c}{GGA} & \multicolumn{1}{c}{$ \text{GGA} + \text{$ p $-fin} $} & \multicolumn{1}{c}{$ \text{GGA} + \text{$ pn $-fin} $} & \multicolumn{1}{c}{$ \text{GGA} + \text{$ pn $-fin} + \text{VP} $} & \multicolumn{1}{c}{All} & \multicolumn{1}{c}{$ \text{All} + \text{ISB} $} \\ \hline
      $ \nuc{He}{4}{} $    & -27.5263 & -27.6120 & -27.6748 & -27.6677 & -27.6597 & -27.6597 & -29.3137 \\
      $ \nuc{O}{14}{} $    & -100.7141 & -100.9292 & -101.5339 & -101.4800 & -101.3917 & -101.3684 & -102.6839 \\
      $ \nuc{O}{16}{} $    & -130.4800 & -130.6925 & -131.2765 & -131.2121 & -131.1245 & -131.1247 & -134.6588 \\
      $ \nuc{O}{24}{} $    & -173.0167 & -173.2203 & -173.7450 & -173.6534 & -173.5690 & -173.5204 & -173.5347 \\
      $ \nuc{Ca}{40}{} $   & -347.0848 & -347.4582 & -349.3631 & -349.1581 & -348.7538 & -348.7544 & -353.5741 \\
      $ \nuc{Ca}{48}{} $   & -415.6148 & -415.9813 & -417.7843 & -417.5394 & -417.1433 & -417.0041 & -417.2803 \\
      $ \nuc{Ni}{48}{} $   & -352.6388 & -353.1148 & -356.0852 & -355.8364 & -355.1244 & -355.3307 & -349.7211 \\
      $ \nuc{Sn}{100}{} $  & -811.6641 & -812.3382 & -817.9244 & -817.3322 & -815.5356 & -815.5896 & -808.9891 \\
      $ \nuc{Sn}{124}{} $  & -1047.2633 & -1047.9111 & -1052.8150 & -1052.1049 & -1050.4031 & -1050.5981 & -1054.9999 \\
      $ \nuc{Sn}{132}{} $  & -1103.0881 & -1103.7325 & -1108.4618 & -1107.7192 & -1106.0438 & -1105.9931 & -1101.0821 \\
      $ \nuc{Sn}{162}{} $  & -1189.5521 & -1190.1618 & -1194.2258 & -1193.4190 & -1191.8476 & -1192.1076 & -1192.7669 \\
      $ \nuc{Pb}{208}{} $  & -1636.6149 & -1637.4850 & -1645.7092 & -1644.4772 & -1640.7825 & -1640.7246 & -1633.4297 \\
      $ \nuc{126}{310}{} $ & -2131.4146 & -2132.5366 & -2145.5436 & -2143.6650 & -2136.3995 & -2136.3397 & -2125.0085 \\
    \end{tabular}
  \end{ruledtabular}
\end{table*}
\begin{table}[tb]
  \centering
  \caption{
    Parameters $ a $ and $ b $ for Eq.~\eqref{eq:sys_Z}.
    For the neutron finite-size effect, Eq.~\eqref{eq:sys_N} is used instead of Eq.~\eqref{eq:sys_Z}.}
  \label{tab:sys_para}
  \begin{ruledtabular}
    \begin{tabular}{ldd}
      & \multicolumn{1}{c}{$ a $ ($ \mathrm{MeV} $)} & \multicolumn{1}{c}{$ b $} \\ \hline
      Direct Coulomb (LDA) & 0.528757 & 1.6692 \\
      Exchange Coulomb (LDA) & -0.390342 & 1.0009 \\
      Exchange Coulomb (GGA) & -0.368013 & 1.0103 \\
      Proton Finite Size & -0.0757012 & 1.0640 \\
      Neutron Finite Size & 0.00706328 & 1.0620 \\
      Vacuum Polarization & 0.00354808 & 1.5765 
    \end{tabular}
  \end{ruledtabular}
\end{table}
\par
All these corrections also change the order of some single-particle levels,
as well as the total energy, 
for example, $ 3s_{1/2} $ and $ 1h_{11/2} $ for protons and $ 1i_{13/2} $ and $ 3p_{1/2} $ for neutrons in $ \nuc{Pb}{208}{} $.
These effects may become even more significant in nuclei with 
stronger Coulomb interaction, such as the proton-rich and the superheavy nuclei.
\par
The corrections to the Coulomb interaction are compared with the ISB energies.
In light nuclei, such as the $ \nuc{O}{16}{} $ case, 
the ISB terms of the nuclear force contribute to the total energy more significantly
than the corrections to the EM interaction.
However, in heavy nuclei, such as the $ \nuc{Pb}{208}{} $ case, 
the contribution of the ISB terms of the nuclear force to the total energy is comparable to the corrections to the EM interaction,
especially, the proton finite-size effect and the vacuum polarization.
Therefore, once the ISB terms are considered, these corrections to the EM interaction should be also considered to keep the consistency.
\par
The charge radii for the selected nuclei are shown in Table \ref{tab:sys_Rch}.
It is seen that
the charge radii calculated in the LDA and those with all the corrections to the Coulomb interaction
are the same at the $ 0.01 \, \mathrm{fm} $ order.
In contrast, the ISB terms of the nuclear interaction can affect the charge radii
around $ 0.1 \, \mathrm{fm} $.
The corresponding radii become smaller, except $ \nuc{Ni}{48}{} $ and $ \nuc{Sn}{100}{} $.
\begin{table}[tb]
  \centering
  \caption{
    Charge radii for the selected nuclei.
    All units are shown in $ \mathrm{fm} $.}
  \label{tab:sys_Rch}
  \begin{ruledtabular}
    \begin{tabular}{rddd}
      \multicolumn{1}{c}{Nuclei} & \multicolumn{1}{c}{LDA}  & \multicolumn{1}{c}{All} & \multicolumn{1}{c}{$ \text{All} + \text{ISB} $} \\ \hline
      $ \nuc{He}{4}{} $    & 2.087 & 2.087 & 2.069 \\
      $ \nuc{O}{14}{} $    & 2.772 & 2.769 & 2.747 \\
      $ \nuc{O}{16}{} $    & 2.771 & 2.768 & 2.758 \\
      $ \nuc{O}{24}{} $    & 2.829 & 2.827 & 2.788 \\
      $ \nuc{Ca}{40}{} $   & 3.486 & 3.482 & 3.475 \\
      $ \nuc{Ca}{48}{} $   & 3.527 & 3.523 & 3.497 \\
      $ \nuc{Ni}{48}{} $   & 3.794 & 3.787 & 3.834 \\
      $ \nuc{Sn}{100}{} $  & 4.509 & 4.504 & 4.511 \\
      $ \nuc{Sn}{124}{} $  & 4.691 & 4.686 & 4.680 \\
      $ \nuc{Sn}{132}{} $  & 4.744 & 4.740 & 4.731 \\      
      $ \nuc{Sn}{162}{} $  & 4.979 & 4.976 & 4.965 \\
      $ \nuc{Pb}{208}{} $  & 5.519 & 5.514 & 5.504 \\
      $ \nuc{126}{310}{} $ & 6.336 & 6.331 & 6.321 \\
    \end{tabular}
  \end{ruledtabular}
\end{table}
\subsection{Mirror nuclei mass difference}
\par
As a test of our framework,
the mirror nuclei mass difference between
$ \nuc{Ca}{48}{} $ and $ \nuc{Ni}{48}{} $
calculated with a combination of the Coulomb LDA or all the corrections
discussed above with the present self-consistent finite-size effects (shown as ``All'' in the table)
and the SAMi or SAMi-ISB functional are shown in Table \ref{tab:mirror}.
Here, in ``(All)'' and ``All,''
the results by the conventional finite-size effects
and the present self-consistent finite-size effects are shown, respectively.
The experimental data are given in AME2016 \cite{Huang2017Chin.Phys.C41_030002}.
\par
At the beginning of this section, we discuss
that the refitting of the functional is not needed unless the results are compared with the experimental data.
Nevertheless, we can compare the calculation results with the experimental data in the mirror nuclei mass difference,
since the contribution of the isospin symmetric part of the functional is basically canceled out.
\par
It is seen that the mirror nuclei mass difference calculated with the Coulomb LDA functional and without the ISB terms of the nuclear force deviate more than $ 4 \, \mathrm{MeV} $ from the experimental data.
Even with the ISB terms of the nuclear force,
still it deviates by more than $ 1 \, \mathrm{MeV} $ although the result is improved.
If all the corrections to the EM contribution are considered on top of the previous finite-size effects,
the error is reduced.
Nevertheless, the error is still around $ 900 \, \mathrm{keV} $.
Once all the corrections with the novel self-consistent finite-size effects are considered in addition to the Coulomb interaction,
the result is further improved and agrees, finally, 
with the experimental data within $ 300 \, \mathrm{keV} $ error.
We should notice that the refit of the SAMi functional may further improve the description of the mirror nuclei mass difference.
\begin{table}[tb]
  \centering
  \caption{
    Mirror nuclei mass difference between $ \nuc{Ca}{48}{} $ and $ \nuc{Ni}{48}{} $ calculated with the combination of the Coulomb LDA or all the corrections to the Coulomb interaction (All) 
    and the SAMi or SAMi-ISB functional.
    Here, ``(All)'' and ``All'' show
    the results by the conventional finite-size effects
    and the present self-consistent finite-size effects, respectively.
    The experimental data given in AME2016 \cite{Huang2017Chin.Phys.C41_030002} are also shown.
    All units are in $ \mathrm{MeV} $.}
  \label{tab:mirror}
  \begin{ruledtabular}
    \begin{tabular}{lddd}
      Functional & \multicolumn{1}{c}{$ \nuc{Ca}{48}{} $} & \multicolumn{1}{c}{$ \nuc{Ni}{48}{} $} & \multicolumn{1}{c}{Difference} \\ \hline
      SAMi \& LDA       & -415.6148 & -352.6388 & 62.9760 \\
      SAMi \& (All)     & -415.7756 & -353.3874 & 62.3882 \\
      SAMi \& All       & -417.0041 & -355.3307 & 61.6734 \\ \hline
      SAMi-ISB \& LDA   & -415.8529 & -347.1168 & 68.7361 \\
      SAMi-ISB \& (All) & -416.0248 & -347.8291 & 68.1957 \\
      SAMi-ISB \& All   & -417.2803 & -349.7211 & 67.5592 \\ \hline
      Exp.~\cite{Huang2017Chin.Phys.C41_030002} & -416.000928 & -348.72 & 67.28 \\
    \end{tabular}
  \end{ruledtabular}
\end{table}
\par
The Nolen-Schiffer anomaly is a related topic to the mirror nuclei mass difference 
\cite{Okamoto1964Phys.Lett.11_150, Nolen1969Annu.Rev.Nucl.Sci.19_471, Saito1994Phys.Lett.B335_17, Shahnas1994Phys.Rev.C50_2346, Meisner2008Eur.Phys.J.A36_37, Menezes2009Eur.Phys.J.A42_97, Dong2018Phys.Rev.C97_021301}.
The anomaly is the difference of mirror nuclei mass difference between theoretical calculation and experimental data.
It is said that this difference comes from both the ISB terms of the nuclear interaction and the Coulomb interaction.
In the present calculation, 
it is, actually, seen that the ISB terms of the nuclear force and the correction to the EM interaction reduce the anomaly.
%
\section{Conclusion and Perspectives}
\label{sec:conc}
\par
In this paper, the finite-size effects of protons and neutrons as well as
the vacuum polarization were considered in a self-consistent Skyrme Hartree-Fock calculation.
The electromagnetic spin-orbit interaction was considered perturbatively.
These contributions to the total energy and their systematic behavior were discussed.
\par
The proton finite-size effect makes the nuclei more strongly bound, for example, for $ 8.2 \, \mathrm{MeV} $ in $ \nuc{Pb}{208}{} $.
In contrast, the neutron finite-size effect makes the nuclei less strongly bound,
and its contribution is almost one order of magnitude smaller than the proton contribution.
The contribution of the vacuum polarization to the total energy is also non-negligible, 
and makes the nuclei less strongly bound,
for example, by $ 3.7 \, \mathrm{MeV} $ in $ \nuc{Pb}{208}{} $.
The contribution of the electromagnetic spin-orbit interaction to the total energy is around $ 50 \, \mathrm{keV} $.
\par
Systematically, the contribution of the isospin symmetry-breaking terms of the nuclear force to the total energy
is comparable to that of the proton finite-size effect in heavy nuclei,
while the former is still more significant than the latter in light nuclei.
The neutron finite-size effect and the vacuum polarization are also non-negligible.
Meanwhile, the contribution of the electromagnetic spin-orbit interaction to the total energy depends on the shell structure.
\par
The mirror nuclei mass difference between $ \nuc{Ca}{48}{} $ and $ \nuc{Ni}{48}{} $ was also calculated.
All the corrections to the Coulomb functional with the SAMi-ISB functional cooperate to reproduce the mirror nuclei mass difference within $ 300 \, \mathrm{keV} $ accuracy,
which is improved from that calculated with conventional finite-size effects.
\par
So far, the spherical symmetry is assumed and the pairing correlations are not considered.
After considering these effects, the systematic study of the mirror nuclei mass difference is promising.
The nuclear structure of the superheavy elements is also an interesting topic for applying the present scheme.
Since the superheavy elements have larger $ Z $,
the highly accurate estimation of the Coulomb contribution to the binding energies is important.
\par
Also, to reach a more accurate estimation of the Coulomb contribution to the binding energy,
study of the Coulomb correlation energy is important,
while this may have certain model dependence because it also includes the effects coming from the nuclear force.
%
\begin{acknowledgments}
  \par
  The authors appreciate Shihang Shen, Enrico Vigezzi, and Kenichi Yoshida for stimulating discussions and valuable comments.
  T.N.~and H.L.~would like to thank the RIKEN iTHEMS program,
  the JSPS-NSFC Bilateral Program for Joint Research Project on Nuclear mass and life for unravelling mysteries of the $ r $ process,
  and the RIKEN Pioneering Project: Evolution of Matter in the Universe.
  TN acknowledges financial support from Computational Science Alliance, the University of Tokyo,
  Universit\`{a} degli Studi di Milano, 
  and the JSPS Grant-in-Aid for JSPS Fellows under Grant No.~19J20543.
  H.L.~acknowledges the JSPS Grant-in-Aid for Early-Career Scientists under Grant No.~18K13549.
  G.C.~and X.R.-M.~acknowledge funding from the European Union's Horizon 2020 research and innovation program under Grant No.~654002.
  The numerical calculations were performed on cluster computers at the RIKEN iTHEMS program.
\end{acknowledgments}
%
\appendix
\section{Isospin Symmetry Breaking Term of SAMi-ISB Functional}
\label{sec:ISB}
\par
In the SAMi-ISB functional \cite{Sagawa1995Phys.Lett.B353_7}, the Skyrme-like zero-range charge-symmetry-breaking (CSB) and charge-independence-breaking (CIB) interactions are 
\begin{align}
  v_{\urm{CSB}} \left( \ve{r}_1, \ve{r}_2 \right)
  & =
    \frac{1}{4}
    \left( \tau_{z1} + \tau_{z2} \right)
    s_0 \left( 1 + y_0 P_{\sigma} \right)
    \delta \left( \ve{r}_1 - \ve{r}_2 \right),
    \label{eq:CSB} \\
  v_{\urm{CIB}} \left( \ve{r}_1, \ve{r}_2 \right)
  & =
    \frac{1}{2}
    \tau_{z1} \tau_{z2}
    u_0 \left( 1 + z_0 P_{\sigma} \right)
    \delta \left( \ve{r}_1 - \ve{r}_2 \right),
    \label{eq:CIB}
\end{align}
respectively, 
where $ \tau_{zi} $ is the $ z $ projection of the isospin operator for the $ i $th nucleon, and
$ P_{\sigma} = \left( 1 + \ve{\sigma}_1 \cdot \ve{\sigma}_2 \right) / 2 $
is the spin projection operator.
The parameters, including the errors attached to each parameter, are
$ y_0 = z_0 = -1 $,
$ s_0 = -26.3 \left( 7 \right) \, \mathrm{MeV} \, \mathrm{fm}^3 $, and
$ u_0 =  25.8 \left( 4 \right) \, \mathrm{MeV} \, \mathrm{fm}^3 $.
For further details on this functional, please see Refs.~\cite{
  Roca-Maza2018Phys.Rev.Lett.120_202501,
  Roca-Maza2018EPJWebConf.194_01002}.
\par
According to these interactions,
the CSB and CIB Skyrme energy densities in the Hartree-Fock calculation are \cite{Roca-Maza2018Phys.Rev.Lett.120_202501}
\begin{align}
  \ca{E}_{\urm{CSB}} \left[ \rho_p, \rho_n \right]
  & =
    \frac{s_0 \left( 1 - y_0 \right)}{8}
    \left(
    \rho_n^2 - \rho_p^2
    \right), \\
  \ca{E}_{\urm{CIB}} \left[ \rho_p, \rho_n \right]
  & =
    \frac{u_0}{8}
    \left[
    \left( 1 - z_0 \right)
    \left(
    \rho_n^2 + \rho_p^2
    \right)
    -
    2 \left( 2 + z_0 \right)
    \rho_n \rho_p
    \right].
\end{align}
Accordingly, the ISB average potentials for protons and neutrons are
\begin{align}
  V_{\urm{ISB}}^p \left( \ve{r} \right)
  = & \,
      \frac{
      u_0 \left( 1 - z_0 \right)
      - 
      s_0 \left( 1 - y_0 \right)}{4}
      \rho_p \left( \ve{r} \right) 
      \notag \\
    & -
      \frac{u_0 \left( 2 + z_0 \right)}{4}
      \rho_n \left( \ve{r} \right)  , \\
  V_{\urm{ISB}}^n \left( \ve{r} \right)
  = & \,
      \frac{
      u_0 \left( 1 - z_0 \right)
      +
      s_0 \left( 1 - y_0 \right)}{4}
      \rho_n \left( \ve{r} \right)
      \notag \\
    & -
      \frac{u_0 \left( 2 + z_0 \right)}{4}
      \rho_p \left( \ve{r} \right)  ,
\end{align}
respectively.
\par
Here, a simple estimation of the ISB energies is discussed.
The proton and neutron distributions are assumed to be the hard sphere as in Sec.~\ref{sec:simple}:
\begin{equation}
  \rho_{\tau} \left( r \right)
  =
  \begin{cases}
    \rho_0^{\tau}
    & r < R_{\tau}, \\
    0
    & r > R_{\tau},
  \end{cases}
\end{equation}
where $ R_{\tau} $ is the radius of the proton or neutron distribution
and
\begin{equation}
  \rho_0^{\tau}
  =
  \frac{3N_{\tau}}{4 \pi R_{\tau}^3}
\end{equation}
is held ($ \tau = p $, $ n $)
with $ N_n = N $ and $ N_p = Z $.
\par
The CSB energy in this simple estimation reads
\begin{align}
  E_{\urm{CSB}}
  & =
    \int
    \ca{E}_{\urm{CSB}} 
    \, d \ve{r}
    \notag \\
  & =
    \frac{s_0 \left( 1 - y_0 \right)}{8}
    \int
    \left[
    \left\{
    \rho_n \left( \ve{r} \right)
    \right\}^2
    -
    \left\{
    \rho_p \left( \ve{r} \right)
    \right\}^2
    \right] 
    \, d \ve{r} 
    \notag \\
  & =
    \frac{s_0 \left( 1 - y_0 \right)}{8}
    \left(
    \frac{4 \pi R_n^3}{3}
    \left( \rho_0^n \right)^2
    -
    \frac{4 \pi R_p^3}{3}
    \left( \rho_0^p \right)^2
    \right)
    \notag \\
  & =
    \frac{s_0 \left( 1 - y_0 \right)}{8}
    \left(
    \frac{3N^2}{4 \pi R_n^3}
    -
    \frac{3Z^2}{4 \pi R_p^3}
    \right)
    \notag \\
  & =
    \frac{s_0 \left( 1 - y_0 \right)}{8}
    \left(
    N \rho_0^n - Z \rho_0^p
    \right),
    \label{eq:CSB_simple}
\end{align}
and the CIB energy reads
\begin{align}
  E_{\urm{CIB}}
  = & \,
      \int
      \ca{E}_{\urm{CIB}} 
      \, d \ve{r}
      \notag \\
  = & \,
      \frac{u_0 \left( 1 - z_0 \right)}{8}
      \int
      \left[
      \left\{
      \rho_n \left( \ve{r} \right)
      \right\}^2
      +
      \left\{
      \rho_p \left( \ve{r} \right)
      \right\}^2
      \right]
      \, d \ve{r}
      \notag \\
    & -
      \frac{u_0 \left( 2 + z_0 \right)}{4}
      \int
      \rho_n \left( \ve{r} \right) \,
      \rho_p \left( \ve{r} \right) \,
      d \ve{r}
      \notag \\
  = & \,
      \frac{u_0 \left( 1 - z_0 \right)}{8}
      \left(
      \frac{4 \pi R_n^3}{3}
      \left( \rho_0^n \right)^2
      +
      \frac{4 \pi R_p^3}{3}
      \left( \rho_0^p \right)^2
      \right)
      \notag \\
    & - 
      \frac{u_0 \left( 2 + z_0 \right)}{4}
      \frac{4 \pi R_p^3}{3}
      \rho_0^n \rho_0^p
      \notag \\
  = & \,
      \frac{u_0 \left( 1 - z_0 \right)}{8}
      \left(
      \frac{3N^2}{4 \pi R_n^3}
      -
      \frac{3Z^2}{4 \pi R_p^3}
      \right)
      \notag \\
    & - 
      \frac{u_0 \left( 2 + z_0 \right)}{4}
      \frac{3NZ}{4 \pi R_n^3}
      \notag \\
  = & \,
      \frac{u_0}{8}
      \left[
      \left( 1 - z_0 \right)
      \left(
      N \rho_0^n + Z \rho_0^p
      \right)
      -
      2 \left( 2 + z_0 \right)
      Z \rho_0^n
      \right].
      \label{eq:CIB_simple}
\end{align}
Here, the usual relationship $ R_n > R_p $ is assumed in Eq.~\eqref{eq:CIB_simple}.
If $ R_n < R_p $ is held, for example, in proton-rich nuclei,
Eq.~\eqref{eq:CIB_simple} is rewritten as
\begin{equation}
  E_{\urm{CIB}}
  = 
  \frac{u_0}{8}
  \left[
    \left( 1 - z_0 \right)
    \left(
      N \rho_0^n + Z \rho_0^p
    \right)
    -
    2 \left( 2 + z_0 \right)
    N \rho_0^p
  \right].
\end{equation}
\par
If $ \rho_0^n = \rho_0^p = \rho_0 / 2 \simeq 0.08 \, \mathrm{fm}^{-3} $ is assumed as in Eq.~\eqref{eq:hard_rhop},
the CSB and CIB energies, Eqs.~\eqref{eq:CSB_simple} and \eqref{eq:CIB_simple}, are evaluated as
\begin{align}
  E_{\urm{CSB}}
  & = 
    \frac{s_0 \left( 1 - y_0 \right)}{8}
    \left( N - Z \right)
    \rho_0
    \notag \\
  & \simeq
    -0.526 \left( N - Z \right)
    \, \mathrm{MeV},
    \label{eq:estimate_CSB} \\
  E_{\urm{CIB}}
  & =
    \frac{u_0}{8}
    \left[ 
    \left( 1 - z_0 \right)
    \left( N + Z \right) \rho_0
    -
    2 \left( 2 + z_0 \right)
    Z \rho_0
    \right]
    \notag \\
  & \simeq
    \left[
    0.516 \left( N + Z \right)
    -
    0.516 Z
    \right]
    \, \mathrm{MeV} 
    \notag \\
  & \simeq
    0.516 N 
    \, \mathrm{MeV}.
    \label{eq:estimate_CIB}
\end{align}
\end{document}